\documentclass[%
superscriptaddress,
longbibliography,
amsmath,amssymb,
prf,]{revtex4-1}
\usepackage[export]{adjustbox}
\usepackage{mathtools}
\usepackage{graphicx}
\usepackage{lineno}
\usepackage{color}
\usepackage{xcolor}
\usepackage{wasysym}
\usepackage{dcolumn}
\usepackage{bm}
\usepackage[colorlinks = true,
            linkcolor = blue,
            urlcolor  = blue,
            citecolor = blue,
            anchorcolor = blue]{hyperref}\usepackage{subfigure}

\newcommand{\xp}{x^{\prime}} 
\newcommand{\yp}{y^{\prime}} 

\begin{document}
\title{Reservoir computing model of two-dimensional turbulent convection}
\author{Sandeep Pandey}
\affiliation{Institute of Thermodynamics and Fluid Mechanics, Technische Universit{\"a}t Ilmenau, D-98684 Ilmenau, Germany.}
\author{J\"org Schumacher}
\affiliation{Institute of Thermodynamics and Fluid Mechanics, Technische Universit{\"a}t Ilmenau, D-98684 Ilmenau, Germany.}
\affiliation{Tandon School of Engineering, New York University, New York, NY 11201, USA.}

\date{\today}

\begin{abstract}
Reservoir computing is an efficient implementation of a recurrent neural network that can describe the evolution of a dynamical system by supervised machine learning without solving the underlying mathematical equations. In this work, reservoir computing is applied to model the large-scale evolution and the resulting low-order turbulence statistics of a two-dimensional turbulent Rayleigh-B\'{e}nard convection flow at a Rayleigh number ${\rm Ra}=10^7$ and a Prandtl number ${\rm Pr}=7$ in an extended spatial domain with an aspect ratio of 6. Our data-driven approach which is based on a long-term direct numerical simulation of the convection flow comprises a two-step procedure. (1) Reduction of the original simulation data by a Proper Orthogonal Decomposition (POD) snapshot analysis and subsequent truncation to the first 150 POD modes which are associated with the largest total energy amplitudes. (2) Setup and optimization of a reservoir computing model to describe the dynamical evolution of these 150 degrees of freedom and thus the large-scale evolution of the convection flow. The quality of the prediction of the reservoir computing model is comprehensively tested by a direct comparison of the results of the original direct numerical simulations and the fields that are reconstructed by means of the POD modes. We find a good agreement of the vertical profiles of mean temperature, mean convective heat flux, and root mean square temperature fluctuations. In addition, we discuss temperature variance spectra and joint probability density functions of the turbulent vertical velocity component and temperature fluctuation the latter of which is essential for the turbulent heat transport across the layer. At the core of the model is the reservoir, a very large sparse random network characterized by the spectral radius of the corresponding adjacency matrix and a few further hyperparameters which are varied to investigate the quality of the prediction. Our work demonstrates that the reservoir computing model is capable to model the large-scale structure and low-order statistics of turbulent convection which can open new avenues for modeling mesoscale convection processes in larger circulation models.
\end{abstract}

\maketitle

\section{Introduction}
\label{sec:1}
The application of machine learning (ML) methods, in particular of deep neural networks (DNN)\cite{Hinton2012,Jordan2015,LeCun2015,Schmidhuber2015,Goodfellow2016}, to fluid flows has transformed the way of processing and analyzing large amounts of data. ML methods are used to parametrize unresolved scales in Reynolds stresses and subgrid scale models for complex physical or geometrical flow configurations at high Reynolds numbers which still remain inaccessible to direct numerical simulations or even large eddy simulations \cite{ling2016reynolds,Kutz2017,Duraisamy2019,fang2019neural,Brunton2020}. They are also used for the detailed segmentation of complex images \cite{Ronneberger2015,fonda2019deep}. DNNs converted large-scale patterns in an extended three-dimensional turbulent convection flow \cite{Emran2015,pandey2018turbulent} into a planar dynamical network where the edges are regions of locally enhanced convective heat flux. Physics-informed and physics-constrained neural networks were developed as a substitute for solving the underlying partial differential equations of the fluid motion \cite{Raissi2019,Geneva2020,Raissi2020}. All supervised ML algorithms make use of the fact that it is often easier to train a DNN with a number of labeled training data of an intended input-output behavior, than to develop a specific numerical code to provide the correct answer for all possible input data. During the training, information propagates forward through the network while weights and biases are updated at each neuron in each hidden layer of the network backwards from the output to the input layer. This back-and-forth iteration (which is called epoch) has to be repeated multiple times until a minimum of the loss function is obtained. Typically, (stochastic) gradient descent algorithms are used for the minimum search of the loss function \cite{Goodfellow2016}. Such a training process requires typically big data records. One further disadvantage of DNNs is their limited capability to process unseen data at very different flow parameters, such as Reynolds or Rayleigh numbers. Possible solutions to overcome this limitation to stick with comprehensive training sets from different cases or to switch to transfer learning methods \cite{Goodfellow2016} which continue the training to adapt a pre-trained DNN to the new input data.
  
Turbulence problems are inherently highly chaotic with stochastic temporal variation of the involved fields. High-dimensional data due to the large grid sizes have to be processed by DNNs to predict statistical properties or the flow evolution; in some cases they fail completely as for example discussed in detail in refs. \cite{King2018,Mohan2019}. These methods suffer from the already mentioned large dimension of the input vector which leads to expensive training procedures of the model. Recurrent neural networks (RNN) are better suited: due to their ``internal memory'' and feedback mechanisms they are by construction more appropriate to learn the dynamics (and thus the resulting statistics of the flow). The long short-term memory (LSTM) network \cite{hochreiter1997long} is a specific subclass of RNNs, which provides a ``gated mechanism'' for information flow in a feedback loop. The internal memory state allows the sequential processing of the data, using a smaller layer size or less layers and processing the time series in smaller batches compared to other DNNs \cite{srinivasan2019predictions}. LSTMs have been applied recently with success for a small Galerkin 9-mode model of a turbulent shear flow \cite{Moehlis_2004} and compared with a standard DNN \cite{srinivasan2019predictions}. In order to circumvent expensive training procedures, echo state networks (ESN) or reservoir computing models (RCM) \cite{jaeger2004ESN,lukovsevivcius2009reservoir} seem to be a further alternative which is considered here. RCM have received recently renewed attention as a method of equation-free modeling of nonlinear dynamical systems, such as for the already mentioned Galerkin 9-mode model of a turbulent shear flow \cite{Pandey2020}, the Lorenz96 model  \cite{vlachas2019forecasting} or the one-dimensional partial differential Kuramoto-Sivashinsky (KS) equation \cite{lu2017reservoir,pathak2018model} on up to 512 grid points. None of the previous examples handles however two-dimensional spatial fields and evaluates reservoir computing on the basis of turbulent statistics. We also mention here that larger numbers of degrees of freedom of these dynamical systems can been processed with parallel versions of RNNs, such as in ref. \cite{vlachas2019forecasting} for a LSTM. 
 
The present work derives a reservoir computing model for a two-dimensional, fully turbulent Rayleigh-B\'{e}nard convection (RBC) flow at a relatively large Rayleigh number of ${\rm Ra}=10^7$ and a Prandtl number of ${\rm Pr}=7$ \cite{chilla2012new,scheel2013resolving} in a domain $\Omega=L\times H$ with an aspect ratio $\Gamma=L/H=6$. Here $L$ is the horizontal length and $H$ the height of the simulation domain. We set up a (Boussinesq-) equation-free model to describe the large-scale dynamics of the flow and to reproduce low-order statistics, such as profiles of the temperature fluctuations and the convective heat flux across the layer as well as the joint statistics of velocity components and temperature. This requires the following subsequent steps: (1) Data-driven reduction of the fully resolved direct numerical simulation (DNS) record to the most energetic degrees of freedom by a standard Proper Orthogonal Decomposition (POD) based on the snapshot method \cite{sirovich1990,park1990,bailon2010,bailon2011} which is applied to the three-dimensional vector field composed of velocity field and temperature fluctuations. This truncation will contain here 83\% of the mean total turbulent energy and cause a compression by 92\%.  (2) Construction and training of a RCM that predicts the dynamical evolution of the most energetic degrees of freedom obtained in step (1) and thus of the large-scale flow and the resulting low-order statistics. We substitute here a Galerkin-truncated reduced-order model, which is obtainable by a projection of the Boussinesq equations of turbulent convection on the individual POD eigenspaces, by the simple dynamics on a large reservoir (which will be explained further below) that does not have any explicit information on the underlying RBC dynamics and is purely data-driven.          

The combined application of POD analysis with a small feedforward neural network on two-dimensional fields was reported by Krischer et al. \cite{Krischer1993}. The usage of RNNs in combination with POD have been reported by Wan et al. \cite{Wan2018} for a two-dimensional Kolmogorov flow, by Vlachas et al. \cite{Vlachas2018} for the Lorenz96 model and the KS equation, by Deng et al. \cite{deng2019time} for the reconstruction of time-resolved turbulent flow measurement from discrete data obtained from the experiments, or by Mohan and Gaitonde \cite{mohan2018deep} for turbulent flow control. Pawar et al. \cite{pawar2019deep} combined POD with a standard DNN for predictions in a heated cavity flow. 

The outline of this paper is as follows. Section \ref{DNS} describes the Boussinesq model of turbulent Rayleigh-B\'{e}nard convection along with numerical procedure and some basic results. Section \ref{POD} illustrates the POD snapshot method. Section \ref{DL} presents the RCM employed in this work and the variation of hyperparameters of the reservoir. Section \ref{results} discusses also the results which are obtained for the present application and provides a comparison with the original DNS and POD based data. We conclude with a summary and an outlook in section \ref{concl}.

\section{Direct numerical simulation}
\label{DNS}
Direct numerical simulations are applied to solve the Boussinesq equations for the two-dimensional case. These equations are made dimensionless by the height of the layer $H$, the free-fall velocity $U_f = \sqrt{g\alpha \Delta T H}$, where $g$ is the acceleration due to gravity, $\alpha$ is the thermal expansion coefficient at constant pressure, and $\Delta T$ represents the imposed temperature difference between the bottom and the top. Times are expressed in units of the free-fall time $T_f = H/U_f$. The Rayleigh number is given by $\mathrm{Ra} = g \alpha \Delta T H^3 / (\nu \kappa)=10^7$, the Prandtl number by $\mathrm{Pr} = \nu/\kappa=7$, and the aspect ratio is $\Gamma=L/H=6$ . Here, $\nu$ is the kinematic viscosity, and $\kappa$ is the thermal diffusivity of the fluid. The equations of motion in dimensionless form, which couple the two-dimensional velocity field $\mathbf{u}=(u_x, u_y)$, the pressure field $p$, and the temperature field $T$, are given by
\begin{align}
\frac{\partial u_x}{\partial x}+\frac{\partial u_y}{\partial y} &=0\,,
\label{eq:1}\\
\frac{\partial u_x}{\partial t}+ u_x \frac{\partial u_x}{\partial x}+u_y \frac{\partial u_x}{\partial y} &=
- \frac{\partial p}{\partial x} + \sqrt{\frac{\mathrm{Pr}}{\mathrm{Ra}}}\left(\frac{\partial^2 u_x}{\partial x^2}+\frac{\partial^2 u_x}{\partial y^2}\right)\,,
\label{eq:2}\\
\frac{\partial u_y}{\partial t}+ u_x \frac{\partial u_y}{\partial x}+u_y \frac{\partial u_y}{\partial y} &=
- \frac{\partial p}{\partial y} + \sqrt{\frac{\mathrm{Pr}}{\mathrm{Ra}}}\left(\frac{\partial^2 u_y}{\partial x^2}+\frac{\partial^2 u_y}{\partial y^2}\right) + T\,,
\label{eq:3}\\
\frac{\partial T}{\partial t}+ u_x \frac{\partial T}{\partial x}+u_y \frac{\partial T}{\partial y} &=
\frac{1}{\sqrt{\mathrm{Pr} \mathrm{Ra}}}\left(\frac{\partial^2 T}{\partial x^2}+\frac{\partial^2 T}{\partial y^2}\right)\,.
\label{eq:4}
\end{align}
No-slip boundary conditions are imposed at the bottom ($y=0$) and top ($y=1$) for the velocity field. The temperature field is constant, $T=1$ at $y=0$ and $T=0$ at $y=1$. The side walls obey periodic boundary conditions for all fields. Equations \eqref{eq:1}--\eqref{eq:4} are numerically solved by the Nek5000 spectral element method package \citep{FISCHER199784}. We use $N_e=48\times 16$ spectral elements and apply Lagrangian interpolations polynomials of order $N=11$ on each element and in each spatial direction. Further details pertaining to the numerical procedure can be found in refs. \citep{scheel2013resolving,pandey2018turbulent}. We sampled the turbulent flow every $0.25 T_f$ for the subsequent POD snapshot analysis.

\begin{figure*}[ht!]
\includegraphics[width=12cm,keepaspectratio]{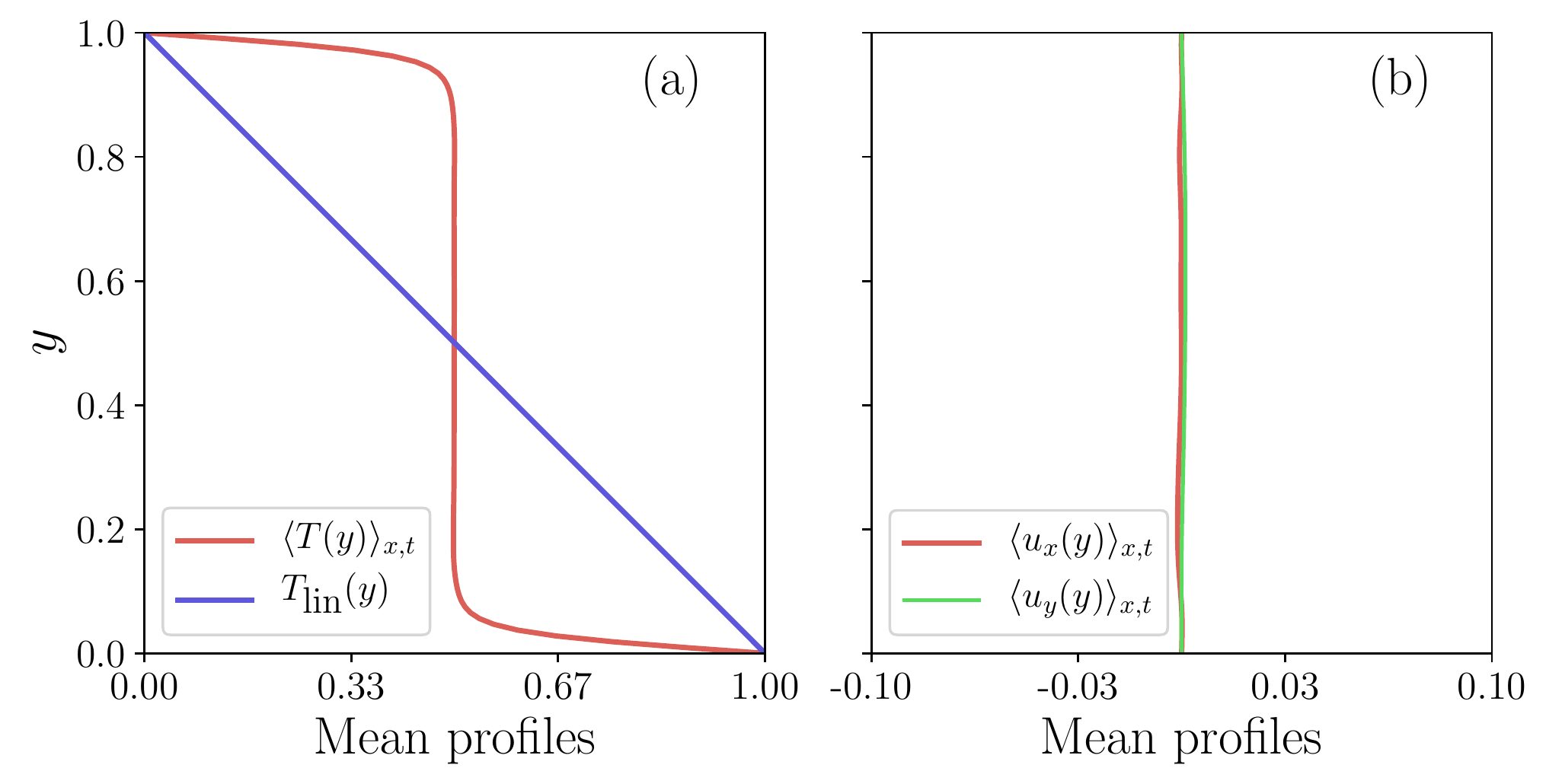}
\caption{Mean profiles of (a): Temperature and (b): Velocity components. The average quantities are obtained by averaging in time and in homogeneous direction $x$. The diffusive equilibrium temperature profile $T_{\rm lin}(y)$ is added to panel (a).}
\label{fig:Fig-1}
\end{figure*}

We apply the standard Reynolds decomposition which is given by
\begin{align}
u_x(x,y,t)&=\langle u_x(y)\rangle_{x,t}+u_x^{\prime}(x,y,t)=u_x^{\prime}(x,y,t)\,,\label{Rey1}\\ 
u_y(x,y,t)&=\langle u_y(y)\rangle_{x,t}+u_y^{\prime}(x,y,t)=u_y^{\prime}(x,y,t)\,,\label{Rey2}\\
T(x,y,t)&=\langle T(y)\rangle_{x,t}+T^{\prime}(x,y,t)\,,
\label{Reynolds}
\end{align}
Mean profiles which have been obtained by a combined $x$-line and time average are displayed in Fig. \ref{fig:Fig-1}. 
In panel (a) of this figure we show linear profile of diffusive heat transfer $T_{\rm lin}(y)=1-y$ which exists for Rayleigh numbers below the instability threshold ${\rm Ra}_c=1708$ together with the mean temperature profile $\langle T(y)\rangle_{x,t}$ for ${\rm Ra}\gg {\rm Ra}_c$. As expected the magnitude of mean velocity is practically zero which is shown in Fig. \ref{fig:Fig-1}(b). This means that $u_x=u_x^{\prime}$ and $u_y=u_y^{\prime}$ as indicated in Eqns. (\ref{Rey1}) and (\ref{Rey2}).
\begin{figure}[ht!]
\centering
\includegraphics[width=15cm]{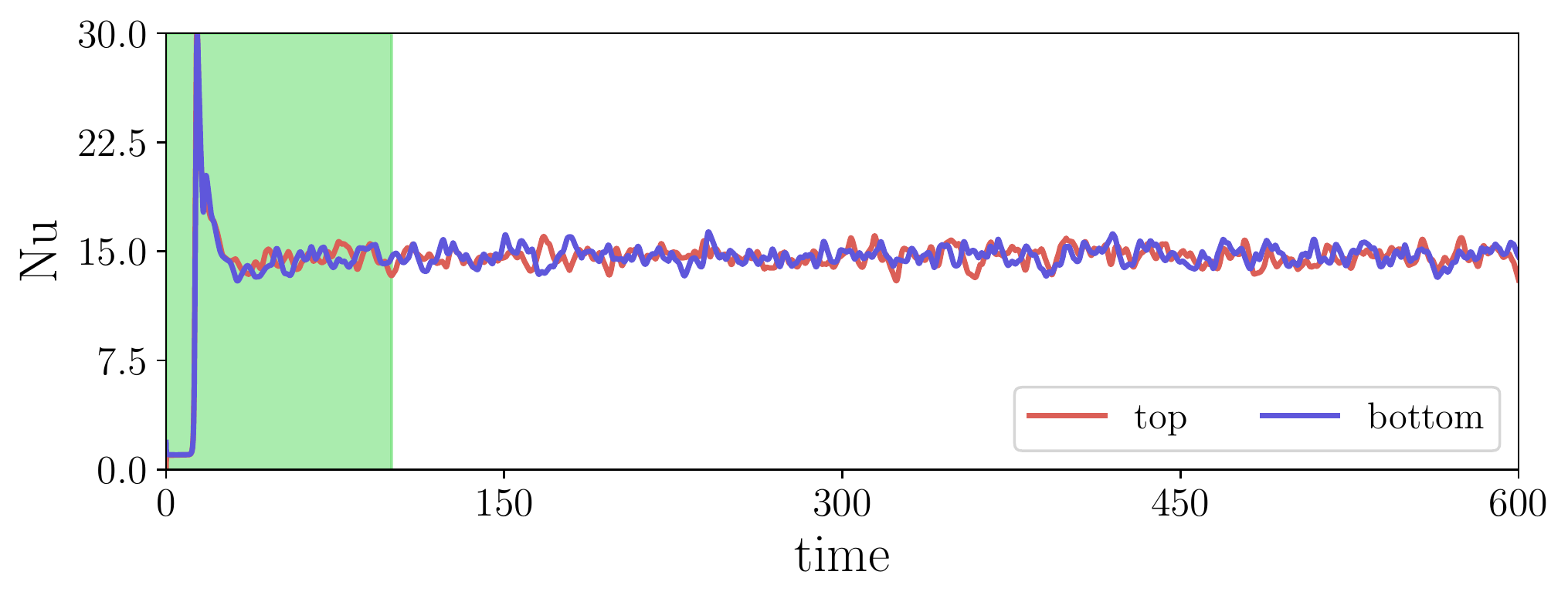}
\caption{\label{A_s_time} Temporal variation of Nusselt number for bottom and top plates obtained in the spectral element DNS. We start sampling our data for times $t>100 T_f$ to exclude the initialization effects which are indicated by the green-shaded region. }
\label{fig:Fig-2}
\end{figure} 
The DNS start from the diffusive equilibrium state and relax into the statistically stationary turbulent convection state after an initial period of $t \approx100 T_f$. Any subsequent snapshot analysis starts for $t > 100T_f$. For the data analysis, we interpolate all DNS snapshots spectrally onto a uniform, somewhat coarser two-dimensional mesh of $N_x\times N_y=160 \times 108$ points. Our data base consists of $N_s=2000$ equidistant simulation snapshots spanning a time range of $500 T_f$. Figure \ref{fig:Fig-2} displays the time series of the Nusselt number, the dimensionless measure of the turbulent heat transfer, which is taken at the bottom and top plates as an average of the diffusive heat flux, ${\rm Nu}(y=0,1)=-\partial\langle T\rangle_{x,t}/\partial y |_{y=0,1}$. The Nusselt number is obtained as an arithmetic average of both means and follows to ${\rm Nu}=14.67 \pm 0.54$. This value agrees well with the results from three-dimensional simulations of RBC for the same parameters which were reported in ref. \cite{fonda2019deep}     

\section{Proper orthogonal decomposition of simulation data}
\label{POD}
Rather than using slices of DNS data as a direct input into a ML algorithm as done for example by King et al. \cite{King2018} in a semi-supervised algorithm, we introduce an intermediate data reduction step. We take a standard POD analysis for simplicity, a prominent method to extract a subset of the energetically dominant degrees of freedom from the fully resolved turbulent convection data \cite{sirovich1990,park1990,bailon2010,bailon2011,podvin2012,podvin2015}. In detail, we will apply the snapshot method \cite{Sirovich1987,weiss2019tutorial} which was developed by Sirovich and Park \cite{sirovich1990,park1990}. Input is the three-dimensional vector field $v_m=(v_1, v_2, v_3)=(u_x, u_y, T^{\prime})$ with zero mean, $\langle v_m \rangle_{x,y,t}=0$. A mean total turbulent energy or variance (which comprises turbulent kinetic energy and scalar fluctuation variance) is given by 
\begin{equation}
E = \Bigg\langle \int_{\Omega} (u_x^2+u_y^2+T^{\prime 2}) d\Omega \Bigg\rangle_t = \Bigg\langle \int_\Omega v_i^2 d\Omega \Bigg\rangle_t\,.
\label{eq:6}
\end{equation}
We want to determine POD modes $\Phi_m(x,y)$ that maximize the averaged projection of $v_m$ onto $\Phi_m$, i.e.,  
\begin{equation}
\frac{\langle |(v_m, \Phi_m)|^2\rangle_t}{(\Phi_m, \Phi_m)} \to \mbox{max}\,,
\label{eq:6a}
\end{equation}
where $(\cdot, \cdot)$ denotes a scalar product on $L_2(\Omega,\mathbb{R}^2)\oplus L_2(\Omega)$ with $u_m=(u_x, u_y)\in L_2(\Omega,\mathbb{R}^2)$ and $T^{\prime}\in L_2(\Omega)$ which induces a norm $\|v_m\|=(v_m, v_m)^{1/2}$. We use the Einstein summation convention and drop the summation symbols for same indices, e.g., $(v_m, v_m)$ is the short notation for $\sum_{m=1}^3 (v_m, v_m)$. Variational calculus translates \eqref{eq:6a} into a search of maxima of a constrained functional $J$ \cite{Holmes1996}, i.e.,  
\begin{equation}
\frac{dJ[\Phi_m+\varepsilon \Psi_m]}{d\varepsilon}\Bigg|_{\varepsilon=0}=0\quad\quad\mbox{with}\quad\quad J[\Phi_m]=\langle |(v_m, \Phi_m)|^2\rangle_t - \lambda ( \|\Phi_m \|^2 -1)\,,
\label{eq:6a2}
\end{equation}
with the Lagrangian multiplier $\lambda$ and $\varepsilon \in \mathbb{R}$.  This leads to the following integral equation
\begin{equation}
\int_\Omega \hat{K}_{mn}(x,y,\xp,\yp) \Phi_n^{(p)}(\xp,\yp) dx^{\prime} dy^{\prime} = 
\int_\Omega \langle v_m(x,y,t) \otimes v_n(\xp,\yp,t)\rangle_t \Phi_n^{(p)}(\xp,\yp) dx^{\prime} dy^{\prime} = \lambda_p \Phi_m^{(p)}(x,y)
\label{eq:6b}
\end{equation}
with the Hermitean, non-negative kernel operator $\hat{K}_{ij}$. The tensor product symbol $\otimes$ is omitted for the rest. Indices $m,n=1,2,3$ and the Einstein summation convection are again used. There is no summation over index $p=1,2,\dots $ which stands for the POD modes that all solve \eqref{eq:6b}. For the present case $p<\infty$;  the kernel operator is a $3N_xN_y\times 3N_xN_y$ matrix with $N_x\times N_y$ the size of the uniform data mesh. The number $3N_xN_y$ corresponds to the (finite) number of degrees of freedom, $N_{\rm dof}$, of the flow, which is approximated on a computational grid and consists of three fields. In addition, we use a Fourier expansion in the homogeneous $x$-direction which reduces the POD analysis to the vertical coordinate $y$ only and uses the periodicity along $x$. Thus \eqref{eq:6b} translates into 
\begin{equation}
\int_0^1 \langle F_k(n_x,y,t) F^{\ast}_l(n_x,y^{\prime},t)\rangle_t \Phi_{l,n_x}^{(p)}(y^{\prime}) dy^{\prime} = \lambda_{p,n_x} \Phi_{k,n_x}^{(p)}(y)
\label{eq:6d}
\end{equation}
with $k,l=1,2,3$ and $n_x\in \mathbb{Z}$ (see \cite{bailon2011} for details). Here, 
\begin{equation}
F_k(n_x,y,t) = \frac{1}{L}\int_{-L/2}^{L/2} v_k(x,y,t)\exp\left(-i \frac{2\pi n_x x}{L}\right) dx \quad\mbox{and}\quad 
\Phi_k^{(p)}(x,y)\to\Phi_{k,n_x}^{(p)}(y) \exp\left(i \frac{2\pi n_x x}{L}\right)
\label{eq:6e}
\end{equation}
Thus the velocity components and temperature fluctuation are expanded in the following POD base
\begin{equation}
v_k(x,y,t) = \sum_{p=1}^{N_{\rm dof}}\,\sum_{n_x=-N_x/2}^{N_x/2} a_{p,n_x}(t)\Phi_{k,n_x}^{(p)} (y) \exp\left(i\frac{2\pi n_x x}{L}\right)\,,
\label{eq:6f}
\end{equation}
with $p=1, \dots, 3N_xN_y=N_{\rm dof}=51840$ (see section II), $k=1,2,3$,  $n_x=-N_x/2, ... N_x/2$, and the reality condition $a_{p,n_x}(t)=a^{\ast}_{p,-n_x}(t)$. Equation \eqref{eq:6b} (or \eqref{eq:6d}) would correspond to a large eigenvalue problem which has to be solved by a direct method. Following Sirovich \cite{Sirovich1987}, the complexity of this task can be reduced. The snapshot method converts the original eigenvalue problem \eqref{eq:6b} for the kernel operator into one for a $N_s\times N_s$ (snapshot) matrix with $N_s\ll N_{\rm dof}$. We therefore chose the following expansion ansatz for each POD mode
\begin{equation}
\Phi^{(p)}_{k,n_x}(y) = \sum_{t_s=1}^{N_s} \beta^{(p)}_{n_x}(t_s) F_k(n_x,y,t_s)\,.
\label{eq:6f}
\end{equation}
We substitute the time average by an arithmetic average over the $N_s$ snapshots. Expansion \eqref{eq:6f} is inserted into \eqref{eq:6d} and results to the following approximation
\begin{equation}
\frac{1}{N_s} \sum_{t_s^{\prime}=1}^{N_s} \int_0^1  F^{\ast}_k(n_x,\yp,t_s) F_k(n_x,\yp,t_s^{\prime}) \beta^{(p)}_{n_x}(t_s^{\prime})\, dy^{\prime} = \lambda_{p,n_x} \beta^{(p)}_{n_x}(t_s)\,. 
\label{eq:6g}
\end{equation}
One has to find the $N_s$ eigenvectors $\beta^{(p)}_{n_x}(t_s)$ and eigenvalues $\lambda_{p,n_x}$ which completes the procedure \cite{bailon2011}. Here, $t_s$ is the integer that stands for DNS snapshot at time $t_s$. Note that the index $p$ runs now from 1 to $N_s$ only. 
\begin{figure}[t]
\centering
\includegraphics[width=14cm,keepaspectratio]{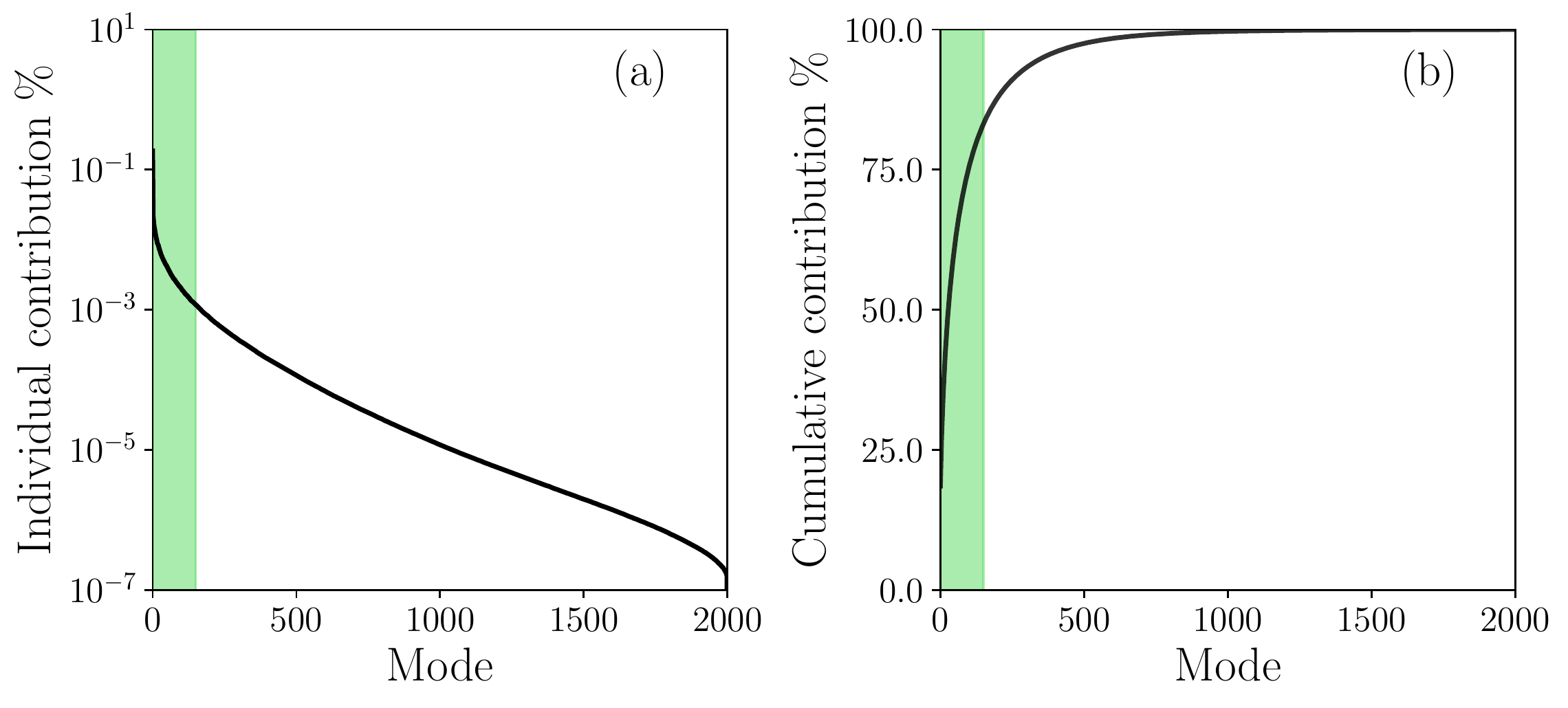}
\caption{\label{POD_Spectra} Eigenvalue spectrum of POD modes as obtained from the analysis of 2000 DNS snapshots. (a): Individual contribution of each mode, and (b): Cumulative contribution of modes. The shaded region show the contribution of first 150 modes which capture the $83\%$ of the total energy of the convection flow and thus provides a good approximation of the large-scale structure.}
\label{fig:Fig-3}
\end{figure} 

\begin{figure*}[ht!]
\centering
\includegraphics[width=15cm,keepaspectratio]{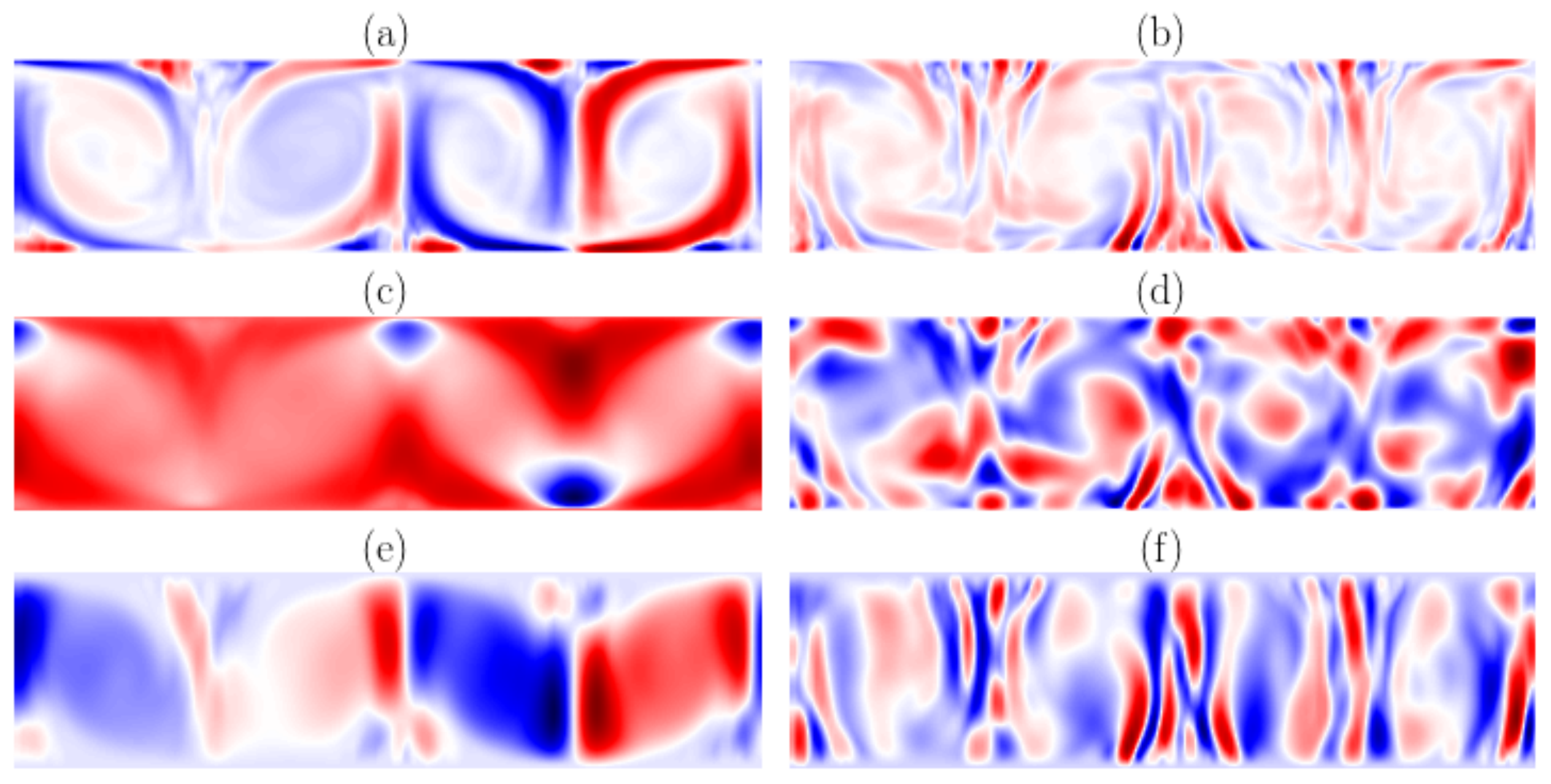}
\caption{Spatial structure of two POD modes. Temperature fluctuation field $T^{\prime}(x,y,t_0)$ snapshot taken at time $t_0$ and resulting by projection onto (a) $\Phi_{3}^{(1)}(x,y)$ and (b) $\Phi_{3}^{(50)}(x,y)$. Horizontal velocity component $u_x(x,y,t_0)$ by projection onto (c) $\Phi_{1}^{(1)}(x,y)$ and (d) $\Phi_{1}^{(50)}(x,y)$. Wallnormal velocity component $u_y(x,y,t_0)$ by projection onto (e) $\Phi_{2}^{(1)}(x,y)$ and (f) $\Phi_{2}^{(50)}(x,y)$. These modes are shown in simulation domain and were obtained by the inverse Fourier transform.}
\label{fig:Fig-4}
\end{figure*}

Figure \ref{fig:Fig-3} displays the spectrum of the eigenvalue analysis. The $N_s$ eigenvalues are sorted with respect to their magnitude which stands for the energy contained in the corresponding POD mode. We denote these modes again by $\Phi_m^{(p)}$. As typical for a turbulent flow, the spectrum falls off quickly, but shows a long tail. We will truncate the POD mode expansion to the $N_{\rm POD}=150$ most energetic POD modes ($N_{\rm POD}<N_s$) which contain 83 \% after the total energy $E$ as given by \eqref{eq:6}. This is shown in Fig. \ref{fig:Fig-3}b. Other truncation levels can be taken, but would not change the subsequent results qualitatively. Figure \ref{fig:Fig-4} illustrates the spatial structure of the three components of the modes $\Phi_m^{(1)}(x,y)$ and $\Phi_m^{(50)}(x,y)$. We obtain the time-dependence of the expansion coefficients $a_{p}(t)$ of the most energetic POD modes $\Phi_m^{(p)}$ by 
\begin{align}
a_p(t) &= \left(v_m(t), \Phi_m^{(p)}\right) = \left(\sum_{q,n_x} a_{q,n_x}(t) \Phi_{m,n_x}^{(q)} \exp\left(i\frac{2\pi n_x x}{L}\right), \sum_{n_x^{\prime}}\Phi_{m,n_x^{\prime}}^{(p)} \exp\left(i\frac{2\pi n_x^{\prime} x}{L}\right)\right) \nonumber\\
&= L \sum_{n_x} a_{p,n_x}(t) \,.
\label{eq:6g}
\end{align}
Here, $m=1,2,3$ and $p,q=1,\dots,N_s$. Orthogonality of the POD modes has been used to obtain the final relation. The time series $a_m(t)$ are used to train the reservoir computing model which is discussed in the following section. They are considered as the ground truth (GT). The first step of the RCM setup is now completed.  

\section{Reservoir computing model}
\label{DL}
\subsection{Architecture and training of reservoir computing model}
\label{Architecture}

\begin{figure}[ht!]
\centering
\includegraphics[width=0.7\linewidth]{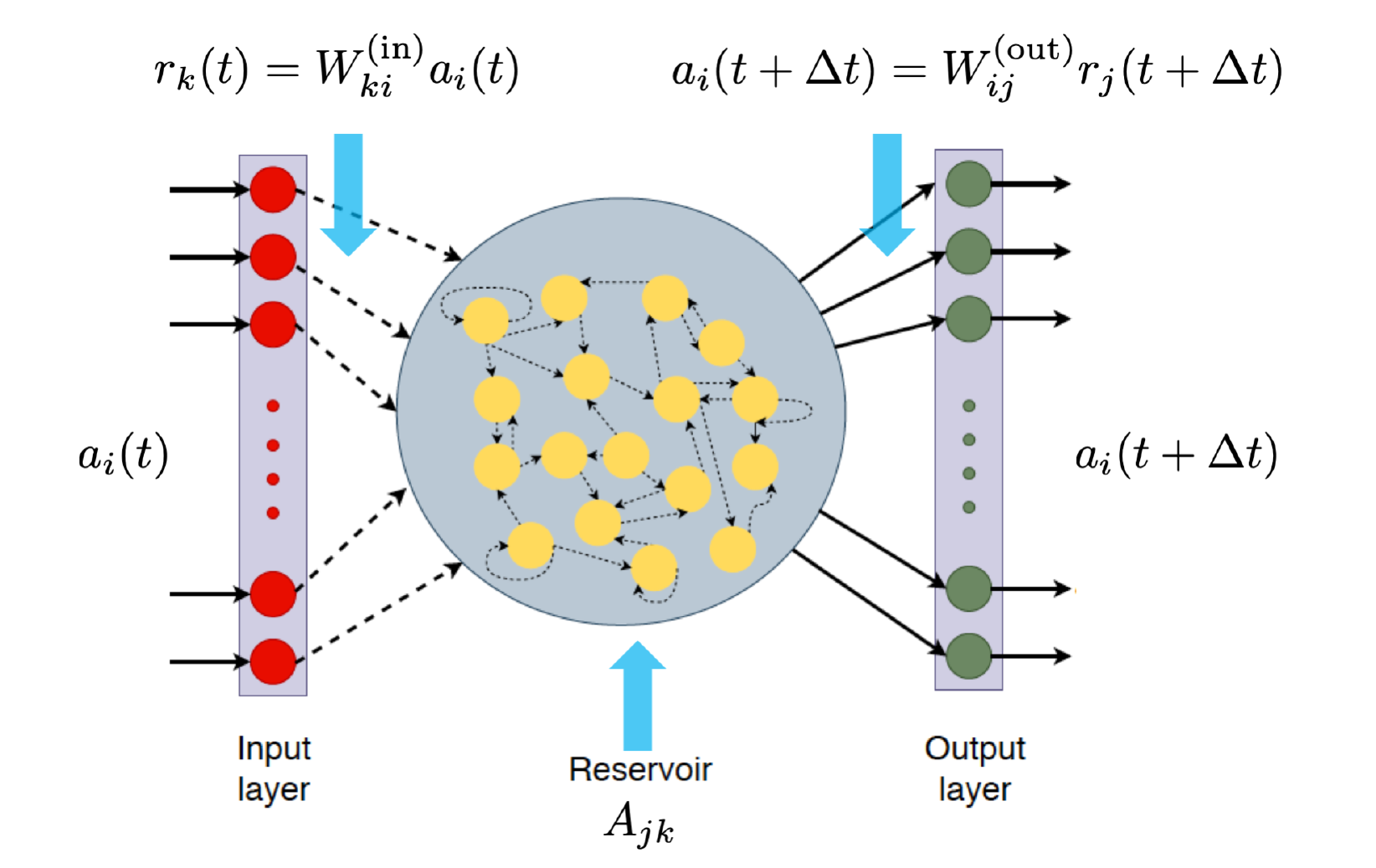}
\caption{The basic architecture of a reservoir computing model RCM (or echo state network ESN) consisting of an input layer, a reservoir, and an output layer. Dotted arrows mark connections which remain fixed during the training. The reservoir consists of $N$ nodes and, both, the input and output layer have  $i=1, \dots, N_{\rm POD}$ units. The reservoir substitutes the stack of hidden layers in a deep neural network.}
\label{fig:Fig-5}
\end{figure}

In the following, we summarize briefly the basics of the RCM, a special type of RNNs \cite{Goodfellow2016}. The RCM architecture is inspired by the brain where many neurons are randomly, recurrently and sparsely connected. Figure \ref{fig:Fig-5} shows the general structure of a RCM. The input consists of training data in form of time series of the POD expansion coefficients $a_i(t)$ with $i=1, \dots, N_{\rm POD}$. They are converted at each instant into a reservoir state vector $r_k(t)\in \mathbb{R}^N$ with $N\gg N_{\rm POD}$. This is done by means of the random weight matrix  $W^{({\rm in})} \in \mathbb{R}^{N\times N_{\rm POD}}$ which is determined at the beginning of the training, 
\begin{equation}
r_k(t) = W_{ki}^{({\rm in})} a_i(t)\,.
\label{RC0}
\end{equation}
Here $N$ is the number of nodes in the reservoir -- a big sparse random network which is described by a (symmetric) adjacency matrix $A \in \mathbb{R}^{N\times N}$. The initialization of $A$ is again random and different strategies for this initialization have been suggested which can improve the results as shown by Strauss et al. \cite{strauss2012design}. Two important parameters of $A$ are the reservoir density $D$ of active nodes and the spectral radius $\rho(A)$, which is set by the largest absolute value of the eigenvalues. Across the reservoir nodes, a simple nonlinear dynamical system evolves which comprises the short-term memory of the network,
\begin{equation}
r_j(t+\Delta t) = (1-\alpha)r_j(t)+\alpha \tanh\left[ A_{jk} r_k(t) + W_{jm}^{({\rm in})} a_m(t)\right]\,,
\label{RC1}
\end{equation}
with $j,k=1,\dots,N$ and $m=1,\dots,N_{\rm POD}$. The nonlinearity enters in form of a typical activation function, here a hyperbolic tangent. A further parameter -- the leakage rate $\alpha$ -- enters the model which blends linear and nonlinear contributions. Optimally, the reservoir should be operated close to an instability which implies a spectral radius $\rho(A)\lesssim 1$. The final element is the random output weight matrix $W^{({\rm out})} \in \mathbb{R}^{N_{\rm POD}\times N}$ which maps the updated reservoir vector back to the POD expansion coefficients,
\begin{equation}
a_i(t+\Delta t) = W_{ij}^{({\rm out})} r_j(t+\Delta t)\,.
\label{RC2}
\end{equation}
This procedure is repeated for $N_d=700$ time steps and all reservoir states are saved. The big advantage of the RCM is that the training is performed with respect to the output layer only. Thus a back propagation procedure that is required in case of DNNs is avoided (see e.g. \cite{lu2017reservoir,pathak2018model} for more details). As a consequence RCM obtains training input from the preceding time step only while other RNNs take a time step sequence from the past. The optimized output weight matrix,  $W^{({\rm out})\ast}$, is obtained by a minimization of a regularized quadratic cost function $C$. A regularization term is added to $C$ to tackle the over-fitting problem \cite{Goodfellow2016}. The function is given by 
\begin{equation}
C \left[W^{({\rm out})} \right]=\sum_{k=1}^{N_d} \Big| W_{ij}^{({\rm out})} r_j(k\Delta t)-a_i(k\Delta t)\Big|^2+\gamma\, \mbox{Tr}\left(W_{ij}^{({\rm out})} W_{jm}^{T({\rm out})}\right)\,.
\label{RC3}
\end{equation}
During this training process, the number of nodes $N$, the reservoir density $D$, the spectral radius $\rho(A)$, the leakage rate $\alpha$, and the prefactor $\gamma>0$ of the regularization term of the quadratic cost function are hyperparameters to tune additionally. Prefactor $\gamma$ in \eqref{RC3} is the ridge regression parameter.  The result of this optimization procedure is the matrix $W^{({\rm out})\ast}$ (as already said) which completes the training.

The prediction mode of the RCM after the training with given POD coefficient time series, $a_i(t)$, as an initial condition is given by
\begin{equation}
r_j(t+\Delta t) = (1-\alpha)r_j(t)+\alpha \tanh\left[ A_{jk} r_k(t) + W_{jk}^{({\rm in})} W_{km}^{({\rm out})\ast} r_m(t)\right]\,.
\label{RC4}
\end{equation}
Now the reservoir dynamics is closed and no further input is required. The dynamics of eq. \eqref{RC4} are used to examine the large-scale evolution and low-order statistics of a two-dimensional turbulent convection flow without using the underlying Boussinesq equations (\ref{eq:1})--(\ref{eq:3}). The translation back to the turbulent fields goes via $W_{km}^{({\rm out})\ast}$ to the $a_p(t)$ and the subsequent reconstruction by the corresponding POD modes $\Phi_k^{(p)}(x,y)$. 

\subsection{Implementation of reservoir computing model}
\label{Architecture}

Previous studies that apply reservoir computing were frequently performed for dynamical systems with a smaller number of degrees of freedom as we discussed in the introduction. Here, we want to take the RCM approach to a new level of complexity by an application to a fully turbulent flow in an extended domain. This will result in additional challenges that start with the architecture and training. The data base consists of the last 1400 (out of the originally 2000) snapshots of the turbulent velocity and temperature fields as reconstructed from the 150 POD modes. These most energetic 150 POD modes account for 83\% of the mean total turbulent energy $E$ (see eq. \ref{eq:6}). By selecting 150 modes, we were able to compress our data by 92\%, while losing 17\% of the information. We also checked that the separate variances of the two-dimensional velocity field and the temperature fluctuations have a similar magnitude.

We use a 50-50\% split of this dataset for training and testing. The current problem is related to time-series forecasting; therefore a random splitting of the data is not considered. The first $N_d=700$ snapshots are used for the training of our RCM and the remaining 700 snapshots are exclusively used for (blind) testing of our framework. As already mentioned earlier, there are few tunable configuration parameters, commonly known as hyperparameters. All these parameters represent the dynamics of the reservoir $A$ and thus affect the RCM \citep{jaeger2004ESN, lukovsevivcius2009reservoir}. These hyperparameters have to be tuned carefully. Bayesian optimization \cite{bergstra2011algorithms, snoek2012practical} is one of the sophisticated methods to find out the optimized hyperparameters in any machine learning task. Here, we applied a simpler grid search as a favorable option which resulted in 100 different hyperparameter settings $(N, \rho(A), \alpha)$. We note here that RCM can become linearly unstable once $\rho(A) > 1$ which sets an upper bound for this parameter. On top of this grid search, the ridge regression parameter $\gamma$ had to be tuned.

We use two measures to determine the quality of the RCM output in relation to GT -- the training data. Note again, that ground truth in our case comprises the reconstructions of the large-scale convection flow as obtained from the first 150 POD modes. Therefore, we monitor the following mean square error (MSE) which is given by 
\begin{equation}
\text{MSE} =  \frac{1}{N_d} \sum_{n=1}^{N_d}  \text{MSE(n)}= \frac{1}{N_d} \sum_{n=1}^{N_d} \frac{1}{N_{\rm POD}} \sum_{j=1}^{N_{\rm POD}} \Big | a_j^{GT}(n) - a_j^{RCM}(n)\Big |^2\,. 
\label{error_definition1}
\end{equation}
where $N_{\rm POD}=150$ and $N_d=700$ for training and test data, respectively.  We are however also interested in the actual flow quantities, such as mean and fluctuation profiles across the convection layer, and thus take furthermore the normalized average relative error (NARE) to ground truth, following \cite{srinivasan2019predictions} in this respect. For example, this error is given for the mean temperature profile by
\begin{equation}
E\left[\langle{T}(y)\rangle_{x,t}\right]=  \frac{1}{2 \max_{y\in [0,1]} (|\langle{T(y)}\rangle_{x,t}^{\rm GT}|)} \int\limits_{0}^{1} \Big| \langle{T(y)}\rangle_{x,t}^{\rm GT}-\langle{T(y)}\rangle_{x,t}^{\rm RCM}\Big| dy  \times 100  \% \,.
\label{error_definition}
\end{equation}
Table \ref{table_reservoir} summarizes the MSE for both, training and test phase, for different values of reservoir node number $N$ (5 different values), spectral radii $\rho(A)$ (5 different values), and leakage rates $\alpha$ (4 different values). In order to obtain the prescribed spectral radius, an adjacency matrix $\tilde A$ with the prescribed reservoir density is initialized randomly first. The spectral radius $\rho(\tilde{A})$ is determined subsequently and the desired spectral radius $\rho(A)$ is enforced afterwards by a rescaling $A=\tilde{A} \rho(A)/\rho(\tilde{A})$. The hyperparameter tuning proceeds as follows: a hyperparameter grid $(N, \rho(A), \alpha)$ with $5\times 5\times 4=100$ tripels is investigated for three different ridge regression parameters $\gamma$. The MSE is determined together with the three NARE for mean temperature, temperature fluctuations, and convective heat flux for each of the 300 cases. The three NARE values follow typically the same trend in the parameter grid search. These NARE magnitudes are calculated after the training or test runs by reconstructing the turbulence fields and performing the combined line-time average as given in \eqref{error_definition}. The tables for the three NARE, which would correspond to table \ref{table_reservoir}, are not shown here but have been evaluated. It is observed that the reservoir dynamics show high sensitivity on all these hyperparameters. The joint evaluation by means of MSE and NARE avoids overfitting. For example, the MSE shows its lowest training value at $\rho(A)= 0.4$ (see columns 3 and 4 of table \ref{table_reservoir}), but the reconstructed flow results in a high NARE amplitudes. This supports our earlier argument regarding the necessity of additional NARE monitoring of flow quantities rather than solely relying on the MSE measure of the POD expansion coefficients.

Table \ref{optimized_para} depicts the final optimal values from the grid search. These values are obtained by a cross-validation procedure, i.e., by monitoring both defined measures \eqref{error_definition1} and \eqref{error_definition}. We model a dynamical system that comprises of $N_{\rm POD}=150$ modes as the degrees of freedom. On the one hand, this requires a large number of nodes in the reservoir $A$; as a rule of thumb $N$ should exceed $N_{\rm POD}$ by an order of magnitude. On the other hand, $N$ should not be chosen too large which can cause an overfitting because irrelevant statistical fluctuations in the training data will be learnt by the model \cite{jaeger2002tutorial}. The leakage rate $\alpha^{\ast}=0.95$ is kept close to 1 which indicates that reservoir evolves slowly \citep{lukovsevivcius2009reservoir, lu2017reservoir}. This value gave the smallest training MSE error as seen in two last columns of table  \ref{table_reservoir}. The spectral radius of adjacency matrix of the reservoir is the central parameter to tune \cite{jaeger2002tutorial}.  A spectral radius close to one, which is eventually taken here, drives the reservoir dynamics close to an instability as discussed in \cite{Goodfellow2016}. The training phases are relatively short and inexpensive when compared to standard DNN or LSTM networks since RCMs do not require back-propagation. With the present data size, it took less than a minute when performed on 24 CPUs.

\begin{table}[h!]
\begin{ruledtabular}
\begin{tabular}{cc  cc  cc}
$N$  &  MSE $\times$10$^{-5}$	& $\rho(A)$  		& MSE $\times$10$^{-5}$ & $\alpha$  		& MSE $\times$10$^{-5}$	 \\ 
        &  Training/Test	&   		&  Training/Test &   		& Training/Test	 \\ \hline 
1000 & 1.5/130.0		& 0.1  	&   0.2/68.1    & 0.1  	&   44.6/68.1    \\
1500 & 1.3/83.7		& 0.4  	&   0.1/71.6    & 0.4  	&   9.0/64.3     \\
2100 & 1.2/90.7		& 0.7  	&   0.4/88.4    & 0.7  	&   2.4/75.4     \\
3000 & 1.3/78.9		& 0.95  &   1.2/90.7    & 0.95  &   1.2/90.7     \\
10000 & 1.6/80.9	& 1.0  	&   1.9/97.8    &   &       \\
\end{tabular}
\end{ruledtabular}
\caption{Performance of the reservoir computing model for different hyperparameters tripels $(N, \rho(A), \alpha)$ quantified by the mean square error (MSE) for training and test phases. The tripels consist of the number of reservoir nodes $N$, the spectral radius $\rho(A)$, and the leakage rate $\alpha$. A total of $5\times 5\times 4=100$ tripels were investigated times 3 different ridge regression parameters $\gamma$. The first two columns display results for $(N, \rho^{\ast}, \alpha^{\ast})$, the second for $(N^{\ast}, \rho(A), \alpha^{\ast})$, and the third two columns for $(N^{\ast}, \rho^{\ast},\alpha)$. Here, $N^{\ast}$ = 2100, $\rho^{\ast}$ = 0.95, and $\alpha^{\ast}$ = 0.95 (see also table 2).}
\label{table_reservoir}
\end{table}
\begin{table}[h!]
\begin{ruledtabular}
\begin{tabular}{ccccccccc}
$N^{\ast}$ 	& $\rho^{\ast}(A)$ 		& $\alpha^{\ast}$ 	&  $D^{\ast}$	& $\gamma^{\ast}$	  & MSE & $E\left[\langle(T^\prime)^2 \rangle\right]$  & $E\left[\langle{T}\rangle\right]$ 		& $E\left[\langle{u_y^\prime T^\prime}\rangle\right]$ \\ 
 	& 		&  	& 			&  & Training/Test & Training/Test  & Training/Test 		& Training/Test \\ \hline 
2100		& 0.95  	&   0.95   	&  0.20  			& 	5$\times$10$^{-2}$		&  1.2$\times$10$^{-5}$  / 9$\times$10$^{-4}$ 	 & 0.0071/0.1774		& 0.22/3.70 	&   1.77/6.22   \\
\end{tabular}
\end{ruledtabular}
\caption{Optimized hyperparameters which result from the whole grid search study (and indicated with asterisk). We list the number of reservoir nodes $N$, the spectral radius $\rho(A)$, the leakage parameter $\alpha$, the reservoir density $D$ (not varied), and the prefactor of the regularization term $\gamma$. Also displayed are the four measures: mean square error MSE, see Eq. \eqref{error_definition1}, and the normalized average relative error (NARE) to ground truth as given by Eq. \eqref{error_definition} for temperature fluctuations, mean temperature, and convective heat flux. Optimal hyperparameters are found with respect to all of these four measures.}
\label{optimized_para}
\end{table}

\section{Prediction of two-dimensional turbulent convection}
\label{results}
\begin{figure}[ht!]
\centering
\includegraphics[width=17cm,keepaspectratio]{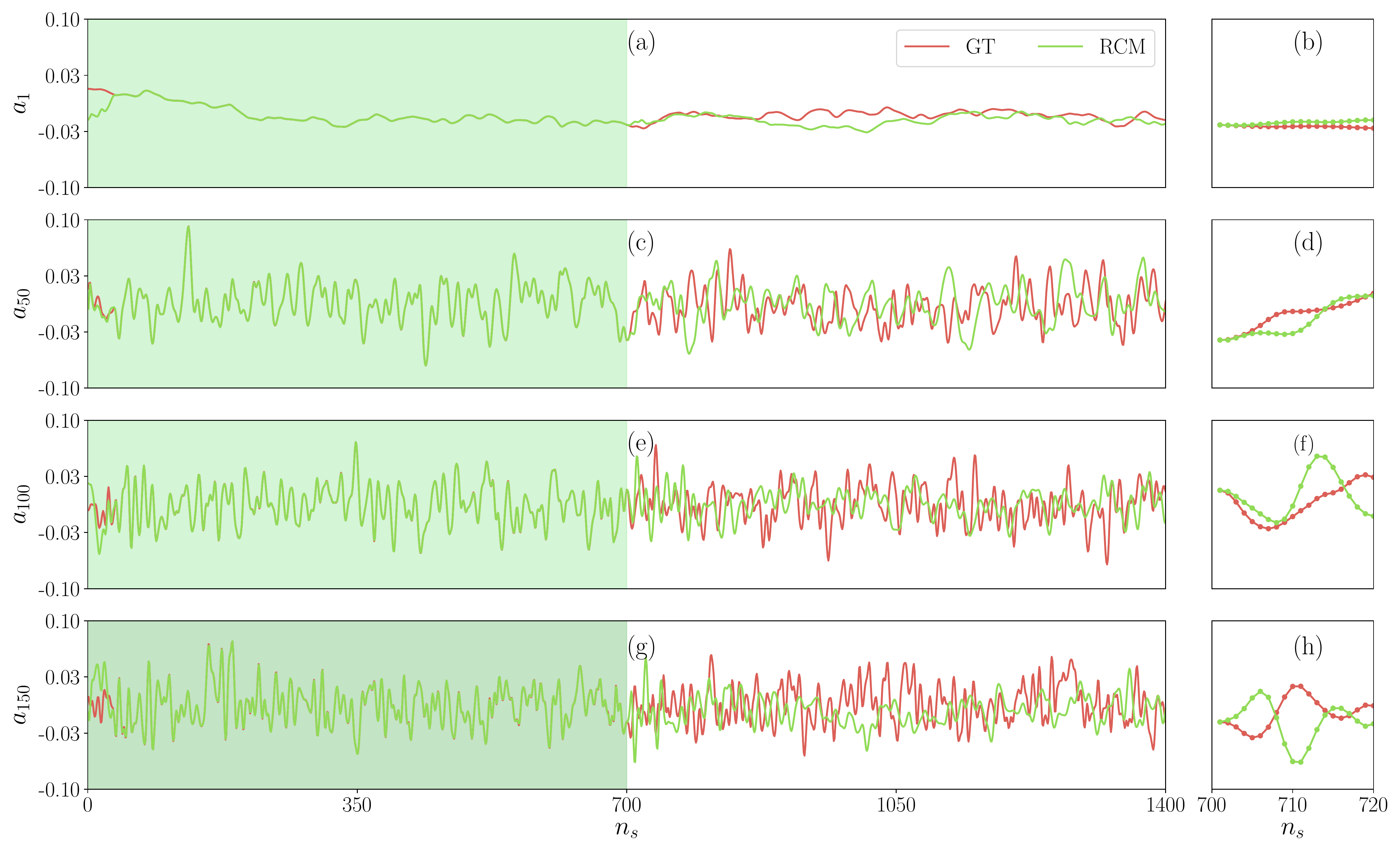}
\caption{Temporal evolution of four POD modes coefficients. From top to bottom: (a,b) $a_{1}(t)$; (c,d) $a_{50}(t)$; (e,f) $a_{100}(t)$, and (g,h) $a_{150}(t)$. The green shaded range shows the training phase while the other range depicts the test phase. Here $n_s=1, \dots, N_s=1400 $ is the snapshot index which translates into the time $t=(n_s/4) T_f$ into a time instant. The corresponding panels in the right column show the initial phase of the forecast. The distance between two snapshots corresponds to 50 integration time steps of the original DNS model.}
\label{fig:Fig-6}
\end{figure} 

Figure \ref{fig:Fig-6}  shows examples of the temporal variation of four selected POD modes --the GT -- together with their predicted time series from our RCM. The predicted time series follow the general trends of the POD data closely, such as amplitudes and frequencies. For instance, the time series of $a_1(t)$ varies on a much smaller frequency as the one for $a_{150}(t)$, which is seen in Figs. \ref{fig:Fig-6}(a,g). A comparison of the exact time evolution of four POD modes is displayed in the panels (b,d,f,h) of Fig. \ref{fig:Fig-6}. The modes with a larger energy content match the GT for a longer period while the modes with a smaller content deviate quickly.  We recall that the turbulent flow modeled here is a highly nonlinear dynamical system. Roundoff errors in the time integration are amplified quickly and lead to an exponentially fast separation of neighboring trajectories in phase space. We conclude that the forecast skills of our RCM model are limited. The GT is a sequence of snapshots which are separated by 0.25 $T_f$ of each other. This corresponds to 50 integration time steps $dt=5\times 10^{-3}$ in the original DNS -- an additional source of roundoff errors.

\begin{figure}[ht!]
\centering
\includegraphics[width=15cm,keepaspectratio]{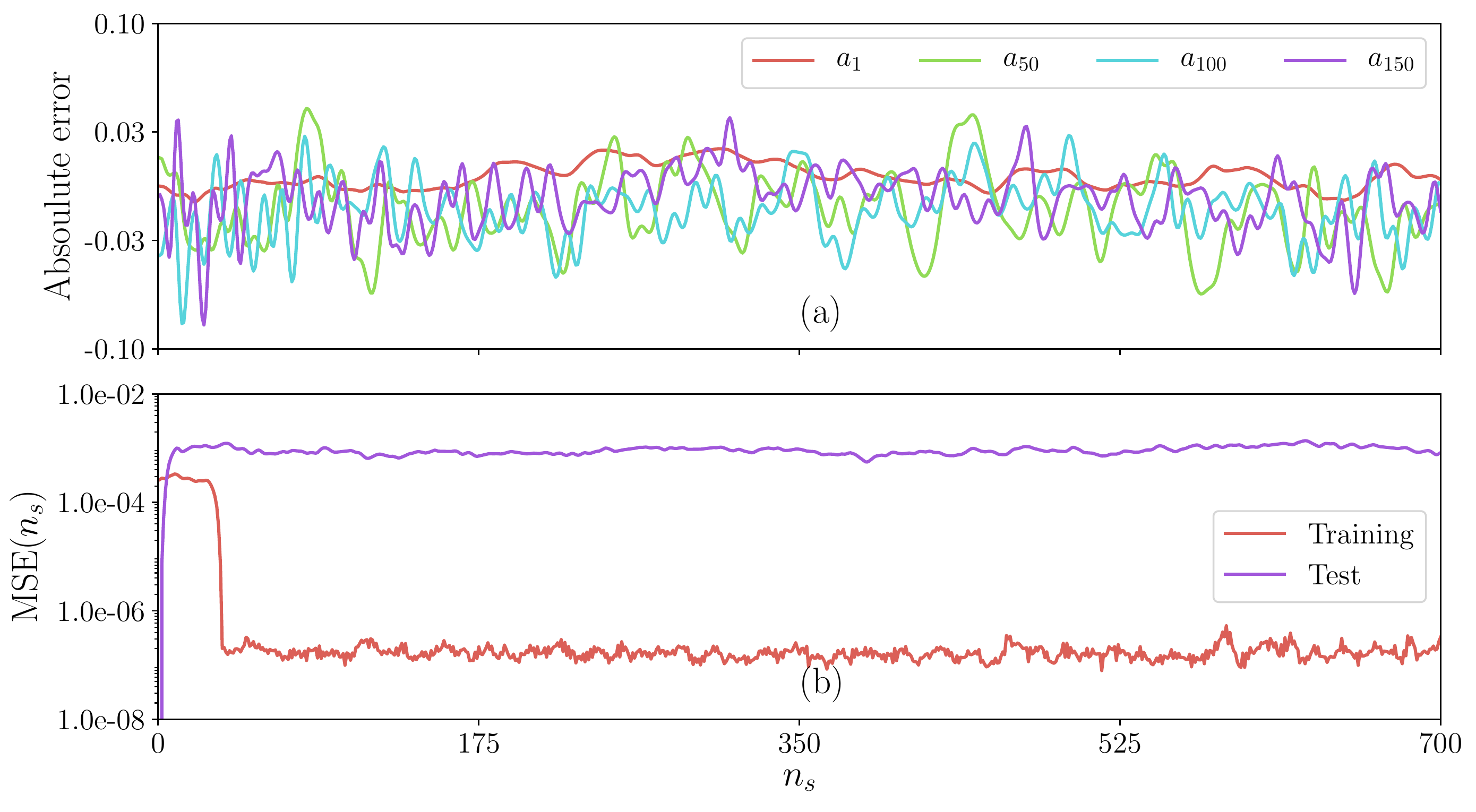}
\caption{Temporal evolution of the error for RCM and GT. (a) Absolute error shown for four individual POD mode coefficients $|a_j^{GT} - a_j^{RCM}|$ with the mode number indicated by the legend. (b) Mean square error MSE($n_s$) versus number of snapshots $n_s$ (see also eq.~\eqref{error_definition1}).}
\label{fig:Fig-6_7}
\end{figure}

Figure \ref {fig:Fig-6_7}a depicts the absolute error of the POD modes coefficients. The error fluctuates around zero mean which suggests that trained model does not suffer from any bias in a specific direction. The MSE as a function of time is also illustrated in Figure \ref {fig:Fig-6_7}b for both training and testing phase. In correspondance with Fig. \ref{fig:Fig-6}, the MSE quickly grows from 0 to 0.0005 with time, but saturates again. The RCM model in its present form is limited in view to a detailed forecast, but can reproduce the low-order statistics as we will see further below. A detailed determination of the forecast quality requires to determine the Lyapunov spectrum of the turbulent flow as done by Pathak et al. \cite{Pathak2017} or Vlachas et al. \cite{vlachas2019forecasting} for other dynamical systems. This is beyond the scope of the present work which focusses on a reproduction of statistical properties.
\begin{figure*}[ht!]
\centering
\includegraphics[width=17cm,keepaspectratio]{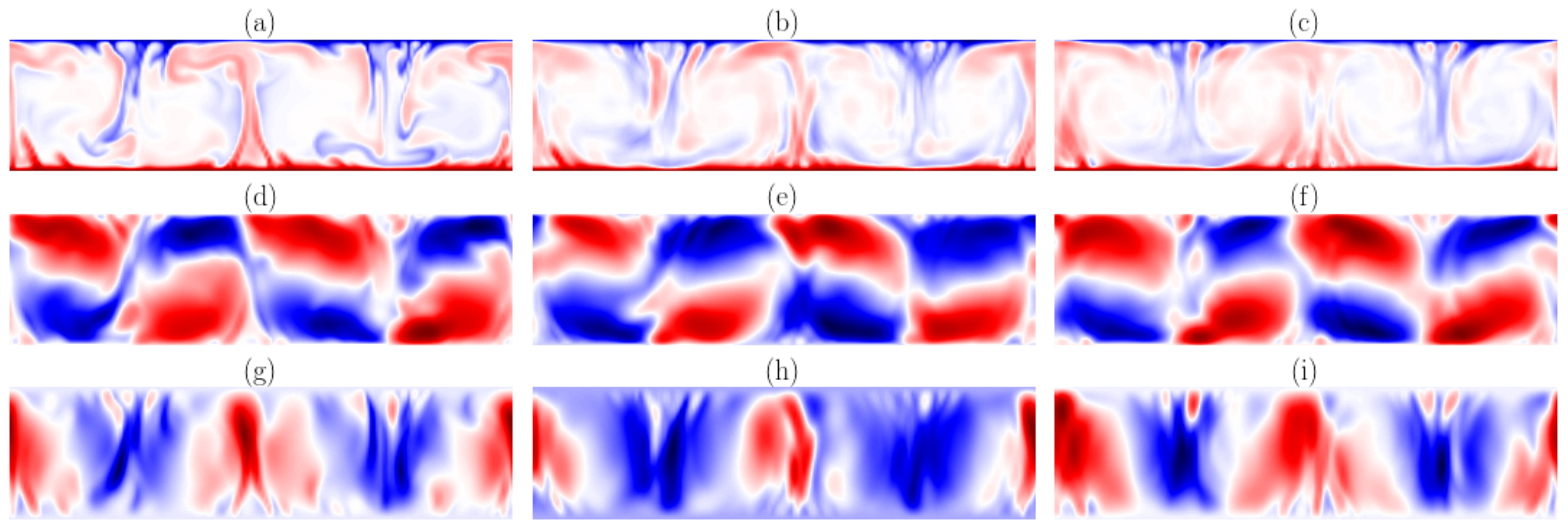}
\caption{Comparison of the turbulence fields in the test phase for $n_s=201$. (a,d,g): Original DNS snapshot. (b,e,h): POD model reconstruction with the 150 most energetic modes which serves as the ground truth for our machine learning problem. (c,f,i): Predicted field from the reservoir computing model. In panels (a,b,c), the instantaneous temperature field $T(x,y,t_0)$ is shown, in (d,e,f) the velocity component $u_x(x,y,t_0)$, and in (g,h,i) the wallnormal velocity component $u_y(x,y,t_0)$.}
\label{fig:Fig-7}
\end{figure*}

Figure \ref{fig:Fig-7} depicts an instantaneous snapshot of all three fields from the original DNS and compares the data with reconstructed field from the 150 POD modes as well as with the prediction from our RCM. We present results only for the blind test data to show the generalization ability of the trained network in predicting all three variables, i.e., the temperature and the two velocity components. It can be observed that the RCM can capture all the large-scale features of the turbulent convection flow with a very good quality even though the forecast skills are limited. This includes a reproduction of the large-scale motion as seen in Figs. \ref{fig:Fig-7}(b,c). Thermal plumes, which are a result of thermal boundary layer instabilities \cite{de2018dynamics}, can also be captured by the RCM which is a challenging task at such high Rayleigh number (see Fig. \ref{fig:Fig-7}a). 

\begin{figure}[ht!]
\centering
\includegraphics[width=16cm,keepaspectratio]{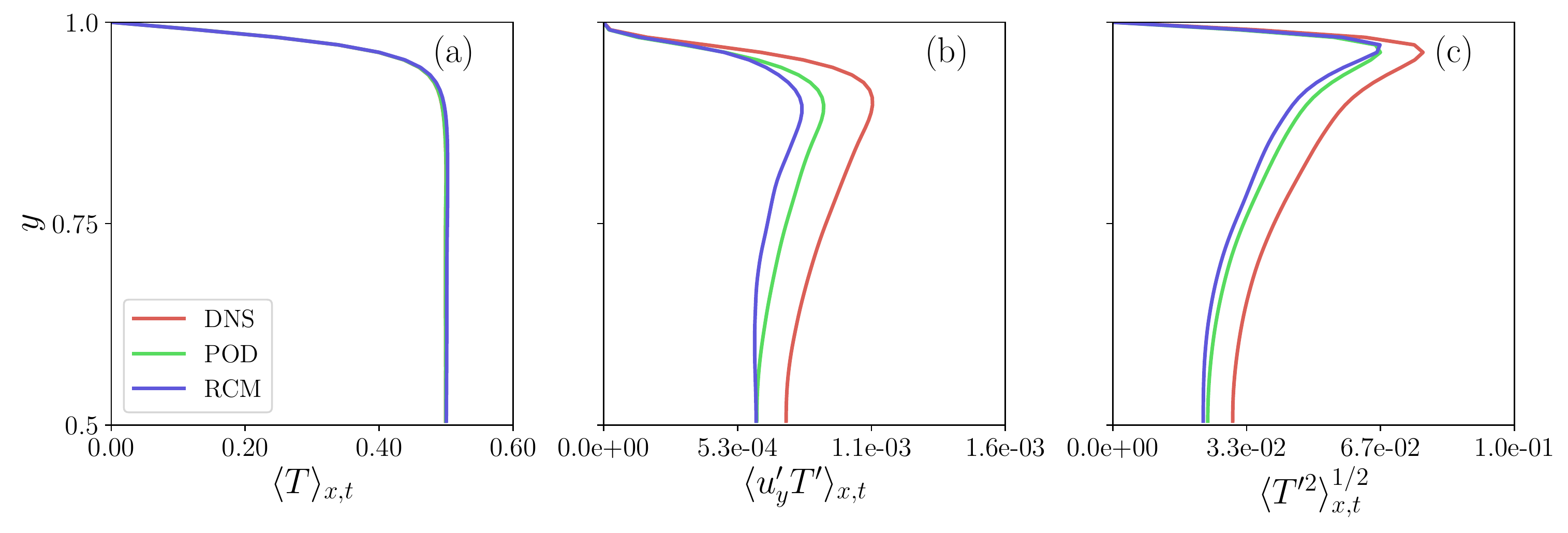}
\caption{\label{A_s_time} Comparison line- and time averaged mean profiles obtained from original DNS, truncated POD, and RCM. (a): Mean temperature, (b): Mean turbulent convective heat flux, and (c): Root mean square fluctuations of temperature. The profiles in case of the RCM  were obtained from the test data set only and thus not used in the training phase.}
\label{fig:Fig-8}
\end{figure}

Next, we report  mean and fluctuations profiles of the flow in Fig. \ref{fig:Fig-8}. Again we show a comparison between the DNS, the projection of the snapshots on the POD modes and the prediction from RCM. In the Boussinesq case, the flow has an up-down-symmetry with respect to the midline at $y = 0.5$ such that we show the upper half only and mirror the data from the lower half. This symmetry which is in line with a transformation $(x,y,u_x,u_y,T^{\prime}) \to (x,1-y,u_x,-u_y,-T^{\prime})$ could possibly be embedded as a physical constraint into the model, a task which we leave for the future work on the three-dimensional case. We also refer here to ref. \cite{mattheakis2019physical} that describes how to embed such a symmetry in a neural network for a Hamiltonian system. 

Figure \ref{fig:Fig-8}a compares the mean temperature profiles $\langle T(y)\rangle_{x,t}$ which is perfectly reproduced by the POD as well as the reservoir computing. Figure \ref{fig:Fig-8}b displays the mean profiles of the turbulent convective heat flux, $\langle u_y^\prime T^\prime \rangle_{x,t}$. It can be seen that the RCM result is in good agreement with the POD data, both close to the wall and the center of the layer. The deviation between the POD data and the DNS data is caused by the truncation at 83\% of the total energy, while the deviation between the POD data and the RCM is caused by the model error.  Figure \ref{fig:Fig-8}c shows the temperature fluctuations $\langle (T^\prime)^2 \rangle^{1/2}_{x,t}$. Again the agreement is very good. To summarize,  the RCM successfully predicts first and second-order statistics as demonstrated by these mean profiles. 

\begin{figure}[ht!]
\centering
\includegraphics[width=16cm,keepaspectratio]{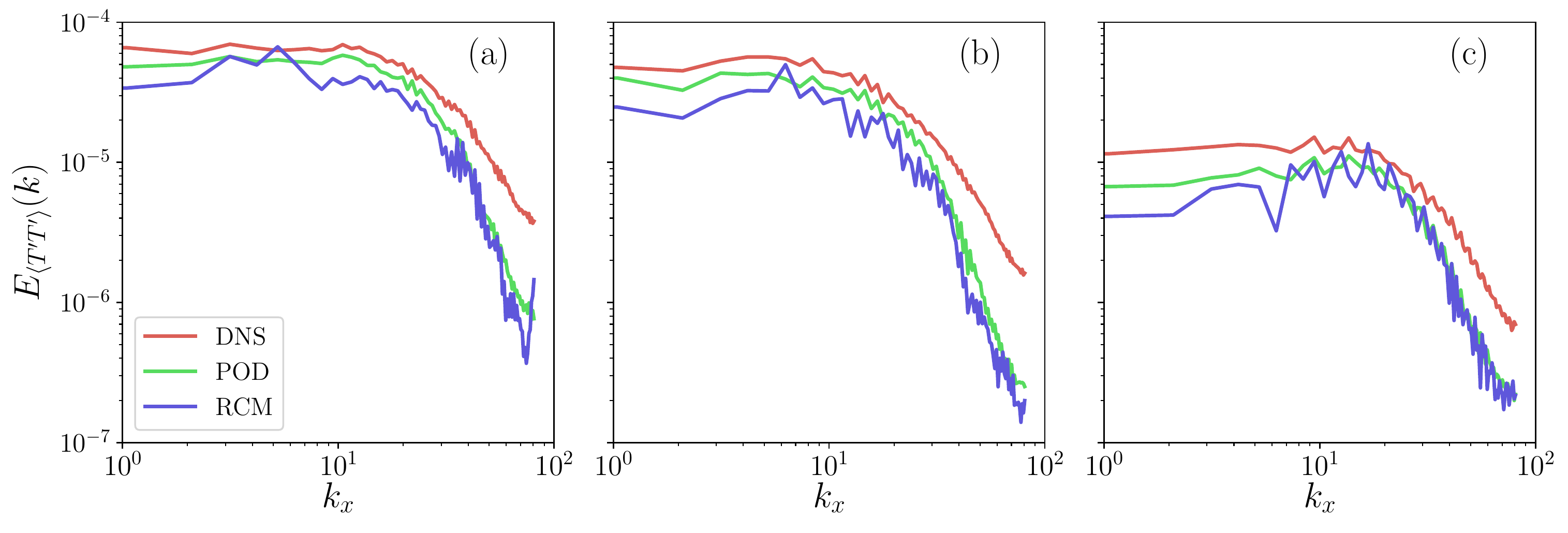}
\caption{Comparison of energy spectra for temperature variance at different distances from the wall (a): $y_0 = 0.07$, (b): $y_0 = 0.15$, and (c): $y_0 = 0.50$. The legend indicates the three data sets taken for the calculation. Line color is the same for all three panels.}
\label{fig:Fig-9}
\end{figure}

To have a thorough comparison, we add results obtained for the Fourier spectra. Figure \ref{fig:Fig-9} depicts the spectrum of the temperature variance at three different locations $y_0$ in the wall-normal direction. As expected, both POD and RCM spectra deviate from the DNS spectra, particularly in the high-wavenumber tail representing the small eddies. The reason is the truncation to 150 POD modes. Our statistics prediction from the RCM  follows however closely the trend of the POD-based data. 

\begin{figure}[ht!]
\centering
\includegraphics[width=16cm,keepaspectratio]{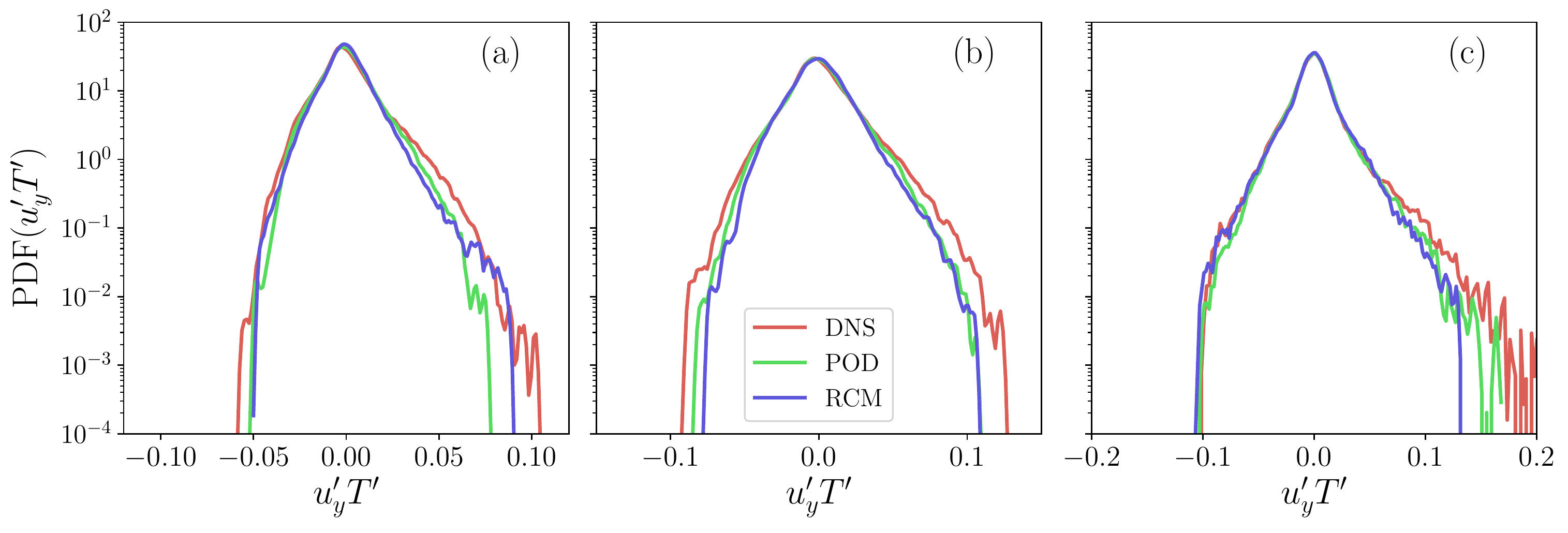}
\caption{Comparison of probability density function (PDF) of the local convective heat flux ($u_y^\prime T^\prime $) at (a): $y = 0.07$, (b): $y = 0.15$, and (c): $y = 0.50$. The legend indicates the three data sets taken for the calculation. Line color is the same for all three panels.}
\label{fig:Fig-10}
\end{figure}

\begin{figure}[ht!]
\centering
\includegraphics[width=16cm,keepaspectratio]{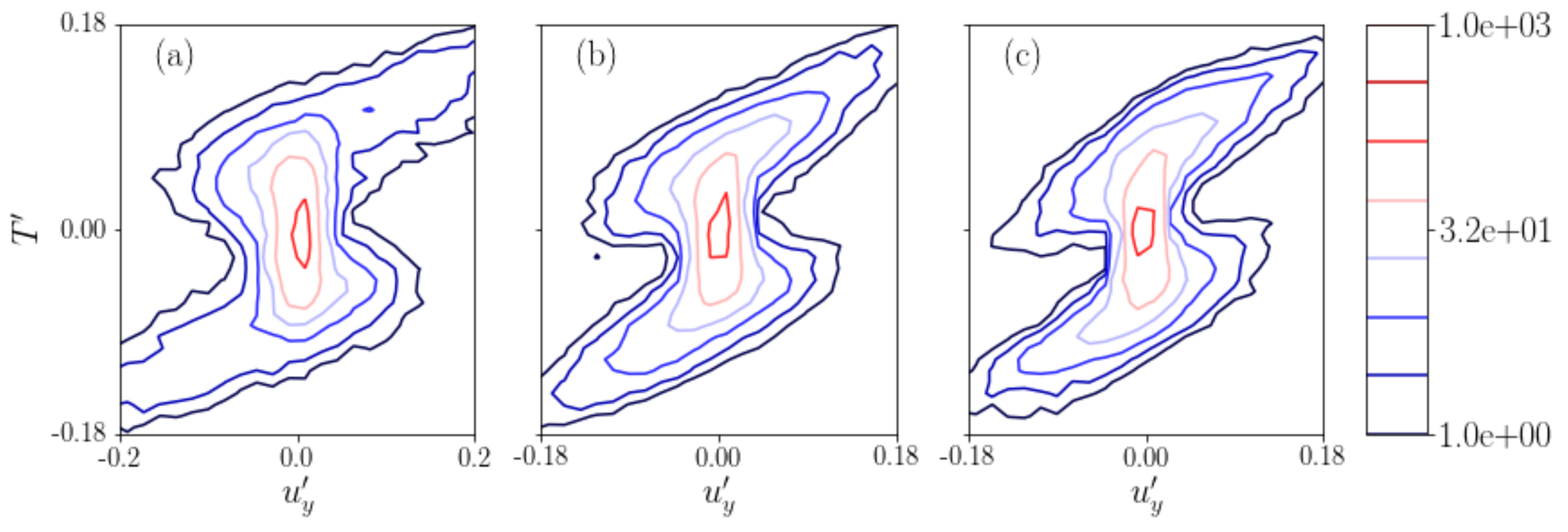}
\caption{Comparison of joint probability density function of the turbulent vertical velocity component and the turbulent temperature fluctuation, $\text{P}(u_y^\prime, T^\prime)$, at the midline at $y = 0.50$ (a): DNS, (b): POD, and (c): RCM. The isocontour levels in the legend are logarithmically spaced and hold for all three panels.}
\label{fig:Fig-11}
\end{figure}

Furthermore, we show a comparison of probability density functions (PDF) of local convective heat flux at three different vertical locations in our turbulent convection domain. Figure \ref{fig:Fig-10} displays the PDFs and we can again observe a very good agreement, even for most of the tails of the distribution. After having found such a good agreement of the PDFs of turbulent heat flux, we assess the joint probability density function of the two individual fluctuating fields of the turbulent heat flux. We have therefore sampled instantaneous fluctuations of the wall normal velocity component $u_y^{\prime}$ and the temperature fluctuation $T^{\prime}$ at the midline. With these data, we determined the joint probability density function and their contours are shown in figure \ref{fig:Fig-11}. The RCM is capable to reproduce the plume ejections from the bottom and the top which correspond to the skewed contours in the first and third quadrants, respectively. 


\section{Conclusions and outlook}
\label{concl}

In the present work, we have discussed a machine learning--based approach to convective turbulence that aimed at a simple modeling of the large-scale evolution and low-order statistics. We applied a specific recurrent neural network with an efficient design, the reservoir computing model also known as the echo state network. Since it is not possible to feed the direct numerical simulation data directly into such a model, in particular in view to future applications to three-dimensional cases, we had to add an intermediate data reduction procedure. In this work, we applied therefore a standard and well-established Proper Orthogonal Decomposition snapshot analysis which extracted the most energetic degrees of freedom in form of empirical orthogonal modes of the fully developed turbulent flow at hand. Turbulence fields which are reconstructed from this finite number of POD modes are the ground truth and serve as training and testing data for the reservoir computing model. The quality of the prediction of the reservoir computing model is in the final part of the work comprehensively tested. This is done by a direct comparison of the RCM results with both, the original direct numerical simulations and the fields reconstructed with the POD modes. In detail, we find a good agreement of the vertical profiles of mean temperature, mean convective heat flux, and root mean square temperature. In addition, we discuss temperature variance spectra and joint probability density functions of the turbulent vertical velocity component and temperature fluctuation, the latter of which is essential for the turbulent heat transport across the layer.

Our presented work should be considered as a first step and a proof-of-concept investigation which certainly can be extended into several directions which we want to discuss briefly now. (1) Although, we report a rather comprehensive analysis of the reservoir parameters, a full Bayesian hyperparameter optimization might improve the performance of the RCM even further. (2) For the present case, it was not necessary to augment the training data since we could generate a long-term DNS record for the two-dimensional case with a reasonable computational effort. In a three-dimensional application this will not be the case anymore as discussed for example in refs. \cite{pandey2018turbulent,fonda2019deep}.    Data augmentation can be done for example by using symmetries of the corresponding turbulent flow, an idea that goes back to Sirovich and Park in their application of POD to convection \cite{sirovich1990,park1990}. Similar aspects of physics-informed ML were recently discussed in another context by Mattheakis et al. \cite{mattheakis2019physical} for the symplectic nature of Hamiltonian systems which were modeled by neural networks. (3) First efforts have been discussed to generalize the RCM or ESN to a deep-ESN which consists of multiple reservoirs for a possible direct simulation data processing \cite{Gallicchio2019}. The successful application of this network architecture to turbulence is yet open.     

To summarize, our presented reservoir computing model could successfully capture essential properties of the dynamics of the larger scales of a turbulent convection flow. It is a recurrent neural network that describes the turbulent convection flow without explicit knowledge of the Boussinesq equations which can lead to interesting applications, e.g., for global circulation models where mesoscale (moist) convection processes  \cite{Pauluis2011} have to be parametrized \cite{Gentine2018}. Our presented efforts will be extended to the three-dimensional RBC case and to subsequent tests of the performance of the RCM with respect to variations of ${\rm Ra}$ and ${\rm Pr}$.      

\begin{acknowledgments}
The work is supported by the Deutsche Forschungsgemeinschaft with Grant No. SCHU 1410/30-1 and in parts by the project ``DeepTurb -- Deep Learning in and of Turbulence" which is funded by the Carl Zeiss Foundation. The generation of the long-term DNS data base was possible by supercomputing resources which were provided by the project grant HIL12 of the John von Neumann Institute for Computing. We thank Christian Cierpka, Florian Heyder, Patrick M\"ader, and Karl Worthmann for fruitful discussions. 
\end{acknowledgments}

\bibliography{references}

\end{document}